\documentclass[nofootinbib,prd,twocolumn,showpacs,showkeys,preprintnumbers]{revtex4}
\usepackage{hyperref,amssymb,amsmath,mathrsfs,bm,graphicx}
\begin{document}
\title{TILTED LEMAITRE--TOLMAN--BONDI SPACETIMES: HYDRODYNAMIC AND THERMODYNAMIC PROPERTIES}
\author{L. Herrera \footnote{Also at U.C.V., Caracas}} 
\email{laherrera@cantv.net.ve}
\affiliation{Departamento   de F\'{\i}sica Te\'orica e Historia de la  Ciencia,
Universidad del Pa\'{\i}s Vasco, Bilbao, Spain}
\author{A. Di Prisco \footnote{Also at U.C.V., Caracas}}
\email{adiprisc@fisica.ciens.ucv.ve}
\affiliation{Departamento   de F\'{\i}sica Te\'orica e Historia de la  Ciencia,
Universidad del Pa\'{\i}s Vasco, Bilbao, Spain}
\author{J. Ib\'a\~nez } 
\email{j.ibanez@ehu.es}
\affiliation{Departamento   de F\'{\i}sica Te\'orica e Historia de la  Ciencia,
Universidad del Pa\'{\i}s Vasco, Bilbao, Spain}
\date{\today}
\begin{abstract}
We consider  Lemaitre--Tolman--Bondi spacetimes from the point of view of a tilted observer, i.e. one with respect to which the fluid is radially moving. The imperfect fluid and the congruence described by its four--velocity, as seen by  the tilted observer is studied in detail. It is shown that from the point of view of such tilted observer the fluid evolves non--reversibly (i.e. with non--vanishing rate of entropy production). The non--geodesic character of the tilted  congruence is related to the non--vanishing of  the divergence of the 4--vector entropy flow. We determine the factor related to  the existence  of energy--density inhomogeneities and describe  its  evolution, these results are compared with those obtained for the non--tilted observer. Finally, we exhibit a peculiar situation where the non--tilted congruence might  be unstable.
\end{abstract}
\date{\today}
\maketitle
\section{Introduction}
It is already a stablished fact that  a variety of line elements may satisfy the Einstein equations for different (physically meaningful) stress--energy tensors (see \cite{1}--\cite{13} and references therein). This ambiguity in the description of the source is generally related to the arbritariness in the choice of the four velocity in terms of which the energy--momentum tensor is split.

Thus, when the two possible interpretations of  a given spacetime correspond to a boost of  one of the observer congruence with respect to the other, both the general properties of the source and the kinematical properties of the congruence  would be different.

This is for example the case of the zero curvature FRW model, which represents a perfect fluid solution for observers at rest with respect to the timelike congruence defined by the eigenvectors of the Ricci tensor, but can also be interpreted as the exact solution for a viscous dissipative fluid as seen by  observers moving relative to the previously mentioned congruence of observers \cite{4}. An important point to mention  is that the relative (``tilting'') velocity between the two congruences may be related to a physical phenomenon such as the observed motion of our galaxy relative to the microwave background radiation.

In other words,   zero curvature FRW models as described by ``tilted'' observers will detect an imperfect fluid, energy--density inhomogeneity, different evolution of the expansion scalar and the shear tensor, among other differences, with respect to the ``standard'' observer (see \cite{4} for a comprehensive discussion on this example). However, as it has been discussed before (e.g see \cite{14}--\cite{16}) imperfect fluids are not necessarily incompatible with reversible processes, accordingly, since the dissipative character of the fluid  defined by the non--vanishing of the divergence of the four--vector entropy flow has an absolute meaning, it is very important to elucidate whether or not entropy production is actually happening.

In the context of the standard Eckart theory \cite{17} a necessary condition for the compatibility of an imperfect fluid with vanishing entropy production (in the absence of bulk viscosity) is  the existence of a conformal Killing vector field CKV) $\chi^\alpha$ such that $\chi^\alpha =\frac{V^\alpha}{T}$ where $V^\alpha$ is the four--velocity of the fluid and $T$ denotes the temperature. In the context of   causal  dissipative  theories, e.g. \cite{18}--\cite{23}, as we shall see below, the existence of such CKV is also necessary for an imperfect fluid to be compatible with vanishing entropy production. 

Now, as it is well known FRW spacetimes admit CKV \cite{24}; therefore tilted FRW even though described by imperfect fluids, might not be associated to irreversible processes (as it seems to be indeed the case, see \cite{15} and \cite{16} for a detailed discussion on this point).

It is our goal in this work to study in detail tilted Lemaitre--Tolman--Bondi spacetimes (LTB). The reasons to undertake such an endeavour are twofold. On the one hand LTB dust models \cite{25, 26, 27} are among the oldest and most interesting solutions to Einstein equations. They describe spherically symmetric distribution of   inhomogeneous non--dissipative dust (see \cite{28, 29} for a detailed description of these spacetimes).

They have been used as cosmological models (see \cite{30, 31, 32, 33, 34} and references therein), in the study of gravitational collapse and the problem of the cosmic censorship \cite{35}--\cite{42}, and in quantum gravity \cite{43, 44}.

A renewed interest in LTB has appeared, in  relation with  recent observations of type Ia supernovae, indicating that the  expansion of the universe is accelerating. Indeed, even if it is true that there is  general  consensus to invoke dark energy as a source of anti-gravity for understanding
the cosmic acceleration, it is also true that a growing number of researchers consider that inhomogeneities can account for the observed cosmic acceleration, without invoking dark energy (see \cite{45, 46, 47, 48, 49, 50, 51, 52} and references therein).

On the other hand, we know that LTB does not admit CKV \cite{53} and therefore heat flux vector appearing in the energy--momentum tensor  of the tilted congruence would be necessarily associated to a ``truly'' (i.e. entropy producing) dissipative phenomenon. Indeed, we shall show  that  in the context of causal dissipative theories too,  the existence of a CKV is necessary for the compatibility of  imperfect fluids with vanishing entropy production.

We shall provide a    possible explanation of such ``truly'' dissipative processes  based on  the non--geodesic character of the tilted congruence. 

The ``inhomogeneity factor'', i.e.  the variable representing those aspects  of the fluid distribution which are responsible for the appearance of energy--density inhomogeneities has been identified for the tilted congruence and  its evolution equation has been integrated. 

Also  a discussion about the stability of the non--tilted congruence and the consequences of attaining the so called ``critical point''  are presented.

Finally a summary of the different issues discussed throughout the text, is given in the last section.

\section{Tilted LTB spacetimes}

We shall now provide a general description of tilted LTB  spacetimes, as well as the basic equations required for our discussion.

Let us consider a line element of the form 
\begin{equation}
ds^2=-dt^2+B^2dr^2+R^2(d\theta^2+\sin^2\theta d\phi^2),
\label{1}
\end{equation}
where $B(r,t)$ and $R(r,t)$ are functions of their arguments, and an energy momentum tensor describing a dust distribution with energy density $\bar \mu$ in comoving coordinates
\begin{equation}
T_{\mu\nu}=\bar \mu v_\mu v_\nu .
\label{Tpo}
\end{equation}

The general form of LTB metric is obtained from the integration of the $(t \ r)$ component of Einstein equations which in turn implies the vanishing of any dissipative flux  in (\ref{Tpo}), the result of such integration produces
\begin{equation}
B(t,r)=\frac{R^\prime}{\left[1+k(r)\right]^{1/2}},\label{BTB}
\end{equation}
where $k$ is an arbitrary function of $r$ and prime denotes derivative with respect to $r$.

In the above we have assumed the congruence to be comoving with the fluid and accordingly
\begin{equation}
v^\mu =(1,0,0,0).
\label{vMc}
\end{equation}

In order to obtain the tilted congruence, let us perform  a Lorentz boost from the Locally Minkowskian frame associated to (\ref{vMc}) to the Locally Minkowskian frame with respect to which a fluid element has radial velocity $\omega$. The corresponding tilted congruence is characterized by the vector field 
\begin{equation}
V^\mu = \left(\frac{1}{(1-\omega^2)^{1/2}},\frac{\omega}{B (1-\omega^2)^{1/2}},0,0\right).
\label{V}
\end{equation}

We shall now assume that the source as seen by the tilted observer consists of an anisotropic fluid (principal stresses unequal) dissipating energy in both, the streaming out limit (a radially directed flow of  a null fluid) and the diffusion approximation (described by means of a heat flow vector $q^\mu$), whose four velocity is given by (\ref{V}). In this case the energy momentum tensor reads:
\begin{eqnarray}
T_{\alpha\beta}=(\mu +
P_{\perp})V_{\alpha}V_{\beta}+P_{\perp}g_{\alpha\beta}+(P_r-P_{\perp})s_{
\alpha}s_{\beta}\nonumber \\+q_{\alpha}V_{\beta}+V_{\alpha}q_{\beta} +\epsilon l_\alpha l_\beta,
\label{tilT}
\end{eqnarray}
where as usual, $\mu, P_r, P_{\perp}$ denote the energy density, the radial pressure and the tangential pressure respectively. $\epsilon$ is the energy density of the null fluid describing dissipation in the streaming out approximation, $l_\alpha$ is a null four vector and the heat flux vector $q^\mu$ satisfying $q^\mu V_\mu=0$, is
\begin{equation}
q^\mu = qs^\mu,
\label{q}
\end{equation}
where
\begin{equation}
s^\mu=\left(\frac{\omega}{(1-\omega^2)^{1/2}},\frac{1}{B (1-\omega^2)^{1/2}},0,0\right),
\label{s}
\end{equation}
and

\begin{equation}
l^\mu=\left(\frac{1+\omega}{(1-\omega^2)^{1/2}},\frac{1+\omega}{B (1-\omega^2)^{1/2}},0,0\right),
\label{l}
\end{equation}
satisfying
\begin{eqnarray}
&&V^{\alpha}V_{\alpha}=-1, \;\; V^{\alpha}q_{\alpha}=0, \;\;
s^{\alpha}s_{\alpha}=1, \;\; s^{\alpha}V_{\alpha}=0,\nonumber\\
&&l_\alpha V^\alpha = -1, \;\; l_\alpha s^\alpha =1, \;\; l_\alpha l^\alpha =0.
\end{eqnarray}
An equivalent form to write the energy momentum tensor is
\begin{equation}
T_{\alpha\beta}=\tilde \mu V_{\alpha}V_{\beta}+\hat{P} h_{\alpha\beta}+\Pi_{\alpha \beta}+\tilde q\left(s_\alpha V_{\beta}+V_{\alpha}s_{\beta}\right), 
\label{tilT}
\end{equation}
where
\begin{equation}
h_{\alpha \beta}=g_{\alpha \beta} + V_\alpha V_\beta, 
\end{equation}
\begin{equation}
\hat P=\frac{\tilde P_{r}+2P_{\bot}}{3},
\end{equation}
\begin{equation}
\Pi= \tilde P_{r}-P_{\bot},
\end{equation}
\begin{equation}
\Pi_{\alpha \beta}=\Pi\left(s_\alpha s_\beta - \frac{1}{3} h_{\alpha \beta}\right),
\end{equation}
\begin{equation}
\tilde \mu=\mu+\epsilon;\;\; \;\;\tilde P_r=P_r + \epsilon; \;\;\;\; \tilde q= q+\epsilon.
\end{equation}
\subsection{Relationships between tilted and non--tilted variables}

From (\ref{Tpo}) and (\ref{tilT}) it is a simple matter to obtain the following relationships linking tilted and non--tilted variables:
\begin{equation}
P_r=\mu-\bar\mu,
\label{til1}
\end{equation}
\begin{equation}
\epsilon=\frac{\bar \mu}{1-\omega^2}-\mu,
\label{til2}
\end{equation}
\begin{equation}
q=-\epsilon-\frac{\bar \mu \omega}{1-\omega^2}=\mu-\frac{\bar \mu}{1-\omega}=\frac{P_r-\mu \omega}{1-\omega}.
\label{til3}
\end{equation}

It is  worth noticing that  for the tilted observer dissipative fluxes ( in either approximation) should be present since $q=\epsilon=0$ implies $\omega=0$.

In the case of bounded configurations the second fundamental form would be discontinuous at the outer boundary implying the presence of a thin shell there.
Some special cases are:
\subsubsection{$\epsilon=0, q\neq 0$}
In this case it follows from the above equations
\begin{equation}
P_r=\mu\omega^2,\qquad \mu=\frac{\bar \mu}{1-\omega^2}, \qquad q=-\frac{\bar\mu \omega}{1-\omega^2}.
\label{til4}
\end{equation}
Observe that the expression for the pressure  corresponds to that of ram pressure, which is intuitively evident.
\subsubsection{$\epsilon\neq 0, q=0$}
In this case it follows from the above equations
\begin{equation}
P_r=\mu \omega, \quad \mu=\frac{\bar \mu}{1-\omega}, \qquad \epsilon=-\frac{\bar\mu \omega}{1-\omega^2}.
\label{til4}
\end{equation}

\subsubsection{$P_r=0$}
In this case we have 
\begin{equation}
\mu=\bar \mu, \qquad q=-\frac{\mu \omega}{1-\omega}, \qquad \epsilon=\frac{\mu \omega^2}{1-\omega^2}.
\label{til5b}
\end{equation}

\subsection{Einstein equations}
In terms of tilted variables,  Einstein's equations $$G_{\alpha \beta} =R_{\alpha \beta} - \frac{1}{2} R g_{\alpha \beta}=8\pi T_{\alpha \beta},$$ take the form
\begin{widetext}
\begin{eqnarray}
8\pi T_{00}=\frac{8\pi}{1-\omega^2}\left(\tilde \mu + \tilde P_r \omega^2 + 2 \tilde q \omega\right)
=\left(2\frac{\dot{B}}{B}+\frac{\dot{R}}{R}\right)\frac{\dot{R}}{R}
-\left(\frac{1}{B}\right)^2\left[2\frac{R^{\prime\prime}}{R}+\left(\frac{R^{\prime}}{R}\right)^2
-2\frac{B^{\prime}}{B}\frac{R^{\prime}}{R}-\left(\frac{B}{R}\right)^2\right],
\label{12} 
\end{eqnarray}
\end{widetext}
\begin{widetext}
\begin{eqnarray}
8\pi T_{01}=-\frac{8\pi B}{1-\omega^2}\left[(\tilde \mu+\tilde P_r) \omega + \tilde q (1+\omega^2)\right] =-2\left(\frac{{\dot
R}^{\prime}}{R} -\frac{\dot B}{B}\frac{R^{\prime}}{R}\right),
\label{13} 
\end{eqnarray}
\end{widetext}
\begin{widetext}
\begin{eqnarray}
8\pi T_{11}=\frac{8\pi B^2}{1-\omega^2}\left(\tilde \mu \omega^2 +\tilde P_r + 2 \tilde q \omega\right) 
=-B^2\left[2\frac{\ddot{R}}{R} + \left(\frac{\dot R}{R}\right)^2\right] 
+\left(\frac{R^{\prime}}{R}\right)^2-\left(\frac{B}{R}\right)^2,
\label{14} 
\end{eqnarray}
\end{widetext}
\begin{widetext}
\begin{eqnarray}
8\pi T_{22}=\frac{8\pi}{\sin^2\theta}T_{33}=8\pi R^2 P_{\perp} 
=-R^2\left(\frac{\ddot{B}}{B}+\frac{\ddot{R}}{R}
+\frac{\dot{B}}{B}\frac{\dot{R}}{R}\right)
+\left(\frac{R}{B}\right)^2\left(
\frac{R^{\prime\prime}}{R}
-\frac{B^{\prime}}{B} \frac{R^{\prime}}{R}\right),\label{15}
\end{eqnarray}
\end{widetext}
where dot denotes derivative with respect to $t$.

Since the Einstein tensor is the same for the tilted and non--tilted observers we should have
\begin{equation}
\bar \mu=\frac{\tilde \mu + \tilde P_r \omega^2 + 2 \tilde q \omega}{1-\omega^2}, \qquad 
(\tilde \mu+\tilde P_r) \omega + \tilde q (1+\omega^2) =0,
\label{nu1}
\end{equation}

\begin{equation}
\tilde \mu \omega^2 +\tilde P_r + 2 \tilde q \omega=0,
\label{nu3}
\end {equation}
which follow at once from (\ref{til1})--(\ref{til3}).  Also  from (\ref{15})
\begin{equation}
P_{\perp}=0.
\label{nu4}
\end{equation}
Fluids with vanishing tangential stresses have been considered in the past in different contexts \cite{54, 55}.

\section{Quantities depending on the congruence}
Since we are going to compare the physical picture as described by two different congruences of observers, it should be obvious that quantities depending explicitly on the congruence would play a fundamental role in such study. We shall consider two different kinds of quantities, namely: kinematical quantities and dynamical quantities, these latter being defined in terms of Riemann and Weyl tensors.
\subsection{Kinematical quantities}
In the absence of rotations (as is the case here) the congruence is described through the three kinematical quantities: the four--acceleration, the expansion and the shear.
The four  acceleration $a^{\alpha}$ is given by
\begin{equation}
a^\alpha=V^\alpha_{;\beta}V^\beta=a s^\alpha,
\label{aal}
\end{equation}
where
\begin{equation}
a=\frac{1}{\sqrt{1-\omega^2}}\left[\frac{\omega \dot B}{B}+\frac{\dot \omega}{ (1-\omega^2)}+\frac{\omega \omega^\prime}{B (1-\omega^2)}\right].
\label{a}
\end{equation}
It is worth noticing that the tilted congruence  is no longer geodesic, in contrast with the non--tilted one. This fact is going to play a relevant role in the physical interpretation of results, as we shall see below.

It is also worth noticing that while a nontrivial solution for $a=0$ might exist for a specific LTB spacetime, producing a specific ``velocity'' field ($\omega$), the generic situation is characterized by $a \neq 0$. At any rate we were unable to find such a solution.

Next, the expansion $\Theta=V^\alpha_{;\alpha}$ is
\begin{equation}
\Theta=\frac{1}{\sqrt{1-\omega^2}}\left[\frac{\dot B}B{}+2\frac{\dot R}{R}+\frac{\omega \dot \omega}{(1-\omega^2)}+\frac{\omega^\prime}{B (1-\omega^2)}+2\frac{\omega R^\prime}{B R}\right].
\label{Th}
\end{equation}

Finally, the shear tensor is defined as usually by 

\begin{equation}
\sigma_{\alpha\beta}=V_{(\alpha
;\beta)}+a_{(\alpha}V_{\beta)}-\frac{1}{3}\Theta h_{\alpha \beta},
\label{Sh}
\end{equation}
which in this particular case may also be written as 

\begin{equation}
\sigma_{\alpha \beta}= \sigma \left(s_\alpha s_\beta -
\frac{1}{3} h_{\alpha \beta}\right), 
\label{sh}
\end{equation}
where
\begin{equation}
\sigma^{\alpha \beta}\sigma_{\alpha \beta} = \frac{2}{3}\sigma^2,
\label{sigsig}
\end{equation}
and
\begin{equation}
\sigma=\frac{1}{\sqrt{1-\omega^2}}\left[\frac{\dot B}{B}-\frac{\dot R}{R}+\frac{\omega \dot \omega}{(1-\omega^2)}+\frac{\omega^\prime}{B (1-\omega^2)}-\frac{\omega R^\prime}{B R} \right].
\label{sig}
\end{equation}

As it is evident from the above, for the non--tilted observer ($\omega=0$) the fluid is geodesic ($a=0$) and the expansion and shear take the  standard form.
\subsection{Dynamical quantities}
Dynamical congruence dependent quantities are defined from the Weyl and Riemann tensors. Thus let us first introduce the Weyl tensor, which is defined through the  Riemann tensor
$R^{\rho}_{\alpha \beta \mu}$, the  Ricci tensor 
$R_{\alpha\beta}$ and the curvature scalar $\cal R$, as:
$$
C^{\rho}_{\alpha \beta \mu}=R^\rho_{\alpha \beta \mu}-\frac{1}{2}
R^\rho_{\beta}g_{\alpha \mu}+\frac{1}{2}R_{\alpha \beta}\delta
^\rho_{\mu}-\frac{1}{2}R_{\alpha \mu}\delta^\rho_\beta$$
\begin{equation}
+\frac{1}{2}R^\rho_\mu g_{\alpha \beta}+\frac{1}{6}{\cal
R}(\delta^\rho_\beta g_{\alpha \mu}-g_{\alpha
\beta}\delta^\rho_\mu). 
\label{Weyl}
\end{equation}
The electric  part of  Weyl tensor (in this particular case the magnetic part vanishes due to the spherical symmetry) is defined by 
\begin{equation}
E_{\alpha \beta} = C_{\alpha \mu \beta \nu} V^\mu V^\nu,
\label{Welec}
\end{equation}
we may also write $E_{\alpha\beta}$ as:
\begin{equation}
E_{\alpha \beta}={\cal E} (s_\alpha
s_\beta-\frac{1}{3}h_{\alpha \beta}),
 \label{We}
\end{equation}
where
\begin{eqnarray}
{\cal E}&= &\frac{1}{2}\left[\frac{\ddot R}{R} - \frac{\ddot B}{B} - \left(\frac{\dot R}{R} - \frac{\dot B}{B}\right)\frac{\dot R}{R}\right]\nonumber \\
&+& \frac{1}{2 B^2} \left[-
\frac{R^{\prime\prime}}{R} + \left(\frac{B^{\prime}}{B} +
\frac{R^{\prime}}{R}\right) \frac{R^{\prime}}{R}\right]
- \frac{1}{2 R^2}. 
\label{E}
\end{eqnarray}
Next, let us introduce the tensors $Y_{\alpha \beta}$, $X_{\alpha \beta}$ and $Z_{\alpha \beta}$  which are elements of the orthogonal splitting of the Riemann tensor and are defined by \cite{56,  57}

\begin{equation}
Y_{\alpha \beta}=R_{\alpha \gamma \beta \delta}V^\gamma V^\delta,
\label{electric}
\end{equation}

\begin{eqnarray}
X_{\alpha \beta}=^*R^{*}_{\alpha \gamma \beta \delta}V^\gamma
V^\delta=\frac{1}{2}\eta_{\alpha\gamma}^{\quad \epsilon
\rho}R^{*}_{\epsilon \rho\beta\delta}V^\gamma V^\delta ,
\label{magnetic}
\end{eqnarray}

 and 
 \begin{equation}
Z_{\alpha \beta}=\frac{1}{2} \eta_{\alpha \gamma \epsilon \rho}R^{\epsilon \rho}_{\;\; \; \; \beta \delta} V^\gamma V^\delta = -\frac{1}{2} \epsilon_{\alpha \epsilon \rho} V^\delta R^{\epsilon \rho}_{\;\; \; \; \beta \delta},
\label{Z}
\end{equation}
where $ \eta_{\alpha \gamma \epsilon \rho}$ is the Levi--Civita tensor,  $\epsilon_{\alpha \epsilon \rho}=V^\gamma \eta_{\gamma \alpha \epsilon \rho}$ and $R^{*}_{\alpha \beta \gamma \delta}=\frac{1}{2}\eta_{\epsilon \rho \gamma \delta} R_{\alpha \beta}^{\quad \epsilon \rho}$.

With these definitions and Einstein's equations we find
\begin{equation}
Y^\beta_\alpha = 4\pi \left(\frac{\tilde \mu}{3} + \hat P \right)h^\beta_\alpha - \left(4\pi \Pi - \cal E\right)\left(s^\beta s_\alpha - \frac{1}{3}h^\beta_\alpha\right),
\label{Ym}
\end{equation}
\begin{equation}
X^\beta_\alpha = 4\pi \left(\frac{2 \tilde \mu}{3} \right)h^\beta_\alpha - \left(4\pi \Pi + \cal E\right)\left(s^\beta s_\alpha - \frac{1}{3}h^\beta_\alpha\right),
\label{Xm}
\end{equation}
\begin{equation}
Z_{\alpha \beta}= - 4 \pi \tilde q^\mu \epsilon_{\alpha \mu \beta}.
\label{Zm}
\end{equation}

Tensors  $Y_{\alpha \beta}$ and $X_{\alpha \beta}$ may be expressed through their traces and their trace-free parts, as
\begin{equation}
Y_{\alpha \beta} = \frac{1}{3}Y_T h_{\alpha \beta} + Y_{TF} \left(s_\alpha s_\beta - \frac{1}{3}h_{\alpha \beta}\right),
\label{Ytatf}
\end{equation}
\begin{equation}
X_{\alpha \beta} = \frac{1}{3}X_T h_{\alpha \beta} + X_{TF} \left(s_\alpha s_\beta - \frac{1}{3}h_{\alpha \beta}\right).
\label{Xtatf}
\end{equation}
From (\ref{Ym}) and (\ref{Xm}) it follows at once that
\begin{equation}
Y_T = 4 \pi  (\tilde \mu+3\hat P), \qquad Y_{TF} = {\cal E} - 4 \pi \Pi,
\label{YTYTF}
\end{equation}
\begin{equation}
X_T = 8 \pi  \tilde \mu, \qquad X_{TF} = -{\cal E} - 4 \pi \Pi.
\label{XTXTF}
\end{equation}
These scalars which obviously are congruence dependent were introduced in \cite{58} and have been  shown to play a relevant role in the study of self--gravitating systems, in particular;
\begin{itemize}
\item  In the absence of dissipation,   $X_{TF}$ controls inhomogeneities in the energy density \cite{58}. 
\item $Y_{TF}$ describes the influence of the local anisotropy of pressure and density inhomogeneity on the Tolman mass \cite{58}
\item $Y_T$ turns out  to be proportional to the  Tolman mass ``density'' for systems in equilibrium or quasi--equilibrium \cite{58}.
\item The evolution of the expansion scalar and the shear tensor  is fully controlled by $Y_{TF}$ and $Y_T$ \cite{53, 58, 59}.
\end{itemize}

Another interesting congruence dependent quantity is the super--Poynting vector $P_\alpha$, which  is associated to any dissipative flux present in the fluid distribution.
From its definition
\begin{equation}
P_\alpha = \epsilon_{\alpha \beta \gamma} \left(Y^\gamma_\delta Z^{\beta \delta} - X^\gamma_\delta Z^{\delta \beta}\right),
\label{spdef}
\end{equation}
and using  (\ref{Ym})--(\ref{Zm}) and (\ref{til1})--(\ref{til3}) we have
\begin{equation}
P_\alpha=32\pi^2 \left(\tilde \mu + \tilde{P_r}\right) \tilde q_\alpha=-32\pi^2 \omega\bar \mu^2 \left[\frac{1+\omega^2}{(1-\omega)^2}\right]s_\alpha.
\label{P}
\end{equation}
Evidently for the non--tilted congruence the super--Poynting vector vanishes as it should be for LTB described by a non--tilted observer.
\section{Some basic auxiliary equations}
For the forthcoming discussion we shall need the explicit form of some basic equations, these are:  the equations of motion , two differential equations relating the Weyl tensor with physical variables and the transport equation for the heat flow.

\subsection{Equations of motion}

The two independent components of Bianchi identities $T^\alpha_{\beta;\alpha}=0$, after some lengthy calculations can be written as 
\begin{eqnarray}
\tilde \mu^\ast&+&\tilde \mu \Theta+\tilde q^\dagger \nonumber \\&+&\tilde q\left(\omega \Theta+\frac{2 R'}{BR}\sqrt{1-\omega^2}+\frac{2\dot \omega}{\sqrt{1-\omega^2}}\right) =0,
\label{Bi0}
\end{eqnarray}

and
\begin{equation}
\tilde P_r^\dagger+(\tilde \mu+\tilde P_r)a +\frac{2\tilde q}{3}\left[2 \Theta+\sigma-3\omega (\ln{R})^\dagger \right]+\tilde q^\ast=0,
\label{Bi1}
\end{equation}
with $f^\dagger=f_{,\alpha}s^\alpha$ and 
$f^\ast=f_{,\alpha}V^\alpha$ .

\subsection{Equations for the Weyl tensor}
As mentioned before  two  differential equations for the Weyl tensor will be needed for the discussion below.  These two equations originally found by Ellis \cite{60, 61} are here reobtained following the procedure adopted in \cite{62} and expressed in terms of $X_{TF}$.
They are
\begin{eqnarray}
(X_{TF}+4\pi \tilde \mu)^\dagger=-3X_{TF}(\ln R)^\dagger\nonumber \\+12\pi\tilde q\left[\omega(\ln R)^\dagger+\frac{\dot R \sqrt{1-\omega^2}}{R}\right],
\label{ellis1}
\end{eqnarray}
and 

\begin{eqnarray}
X_{TF}^\ast=-3X_{TF} (\ln R)^\ast+4\pi \tilde q^\dagger+\frac{4\pi \bar \mu \sigma}{1-\omega^2}\nonumber \\+4\pi\tilde q\left(\frac{2 \dot \omega}{\sqrt{1-\omega^2}}+\omega \Theta-\frac{R'\sqrt{1-\omega^2}}{BR}\right),
\label{ellis2}
\end{eqnarray}
where (\ref{Bi0}) has been used.

\subsection{Transport equation}
 In the diffusion approximation ($\epsilon=0, \tilde q=q$), we shall need a transport equation derived from  a causal  dissipative theory ( e.g. the
M\"{u}ller-Israel-Stewart second
order phenomenological theory for dissipative fluids \cite{18, 19, 20, 21}).

Indeed,  the Maxwell-Fourier law for
radiation flux leads to a parabolic equation (diffusion equation)
which predicts propagation of perturbations with infinite speed
(see \cite{22}, \cite{23}, \cite{63}-\cite{65} and references therein). This simple fact
is at the origin of the pathologies \cite{66} found in the
approaches of Eckart \cite{17} and Landau \cite{67} for
relativistic dissipative processes. To overcome such difficulties,
various relativistic
theories with non-vanishing relaxation times have been proposed in
the past \cite{18, 19, 20, 21, 68, 69}. The important point is that
all these theories provide a heat transport equation which is not
of Maxwell-Fourier type but of Cattaneo type \cite{70}, leading
thereby to a hyperbolic equation for the propagation of thermal
perturbations.

A fundamental parameter   in these theories is the relaxation time $\tau$ of the
corresponding  dissipative process. This positive--definite quantity has a
distinct physical meaning, namely the time taken by the system to return
spontaneously to the steady state (whether of thermodynamic equilibrium or
not) after it has been suddenly removed from it. 
Therefore, when studying transient regimes, i.e., the evolution between two
steady--state situations,  $\tau$ cannot be neglected. In 
fact, leaving aside that parabolic theories are necessarily non--causal,
it is obvious that whenever the time scale of the problem under
consideration becomes of the order of (or smaller) than the relaxation time,
the latter cannot be ignored, since 
neglecting the relaxation time ammounts -in this situation- to
disregarding the whole problem under consideration.

Sometimes in the past it has been argued that dissipative processes with relaxation times
comparable  to the characteristic time of the system are out of the
hydrodynamic regime.  However,  that argument  can be valid only if the particles making up the
fluid are the same ones that transport the heat. But, this is 
never the case. Specifically, for a neutron star, $\tau$ is of the
order of the scattering time between electrons (which carry the
heat) but this fact is not an obstacle (no matter how large the
mean free path of these electrons may be) to consider the neutron
star as formed by a Fermi fluid of degenerate neutrons. The same
is true for the second sound in superfluid Helium and solids, and
for almost any ordinary fluid. In brief, the hydrodynamic regime
refers to fluid particles that not necessarily (and as a matter of fact,
almost never) transport the heat. Therefore large relaxation times (large
mean free paths of particles involved in heat transport) does not imply a
departure from the hydrodynamic regime (this fact has been streseed before
\cite{71}, but it is usually overlooked).

Thus, the transport equation for the heat flux reads
\begin{equation}
\tau h^{\alpha \beta}V^\gamma q_{\beta;\gamma}+q^\alpha=-\kappa h^{\alpha \beta}\left(T_{,\beta}+Ta_\beta\right)-\frac{1}{2}\kappa T^2 \left(\frac{\tau V^\beta}{\kappa T^2}\right)_{;\beta} q^\alpha,
\label{tre}
\end{equation}
where $\kappa$ denotes the thermal conductivity, and $T$ and $\tau$ denote temperature and relaxation time respectively. The transport equation has only one independent component which may be written as
\begin{widetext}
\begin{eqnarray}
\tau \left(\dot q + \frac{\omega q^\prime}{B}\right)+ q(1-\omega^2)^{1/2}&=&
-\kappa\left[\left(\omega \dot T + \frac{T^\prime}{B}\right)+\omega T\left(\frac{\dot B}{B}+\frac{\dot \omega}{\omega(1-\omega^2)}+\frac{\omega^\prime}{B(1-\omega^2)}\right)\right]\nonumber\\
&-&\frac{\kappa T^2}{2}q\left[\left(\frac{\tau}{\kappa T^2}\right)^{\dot{}}+\frac{\omega}{B}\left(\frac{\tau}{\kappa T^2}\right)^\prime\right]- \frac{\tau \Theta}{2}q  (1-\omega^2)^{1/2}.
\label{treq1}
\end{eqnarray}
\end{widetext}
We are now in capacity to analyze some relevant aspects of tilted LTB spacetimes.

\section{Does the tilted observer detect a real dissipative process? (and why?)}
We have seen that for the tilted observer  the energy momentum tensor corresponds to that of an imperfect fluid. However we  know that non--dissipative (reversible) processes within imperfect fluids are not forbidden a priori, in the context of the standard irreversible thermodynamics (see \cite{14, 15, 16} and references therein).

What is the situation for a causal  thermodynamic theory such as Israel--Stewart?
In order to eilucidate this point we shall for simplicity consider in the dissipative part of the energy--momentum tensor only the terms associated to the heat flux, excluding  any shear and bulk viscosity term as well as viscous/heat coupling constants. Then from the Gibbs equation and Bianchi identities it follows (see eq.(42) in \cite{72}) 
\begin{widetext}
\begin{eqnarray}
T S^{\alpha}_{;\alpha} =  
- q^{\alpha} \left[ h^\mu_{\alpha} (\ln{T })_{,\mu} + 
V_{\alpha;\mu} V^\mu
 + \beta_{1} q_{\alpha;\mu} V^\mu+
\frac{T}{2} \left(
\frac{\beta_{1}}{T}V^{\mu}\right)_{;\mu}q_{\alpha}\right],
\label{diventropia}
\end{eqnarray}\\
\end{widetext}
where $S^\alpha$ is the entropy four--current and $\beta_1=\frac{\tau}{\kappa T} $.

Let us first review the situation for the standard irreversible thermodynamics, in this case we have $\tau=0$ and (\ref{diventropia}) becomes 
\begin{eqnarray}
T S^{\alpha}_{;\alpha} =  
- q^{\alpha} \left[ h^\mu_{\alpha} (\ln{T })_{,\mu} +
V_{\alpha;\mu} V^\mu
 \right],
\label{diventropiaI}
\end{eqnarray}
which after simple manipulations takes the form
\begin{equation}
S^{\alpha}_{;\alpha} = -\frac{1}{2} T^{\alpha \beta}_{dis.}
\mathcal{L}_\chi g_{\alpha \beta},
\label{diventropiaII}
\end{equation}
where $\mathcal{L}_\chi$ denotes the Lie derivative with respect to the vector field $\chi^\alpha=\frac{V^\alpha}{T}$, and  $T^{\alpha \beta}_{dis.}=V^\alpha q^\beta+V^\beta q^\alpha$. From the above is evident that if  $\chi$ defines  a conformal Killing vector (CKV), i.e.
\begin{equation}
\mathcal{L}_\chi g_{\alpha \beta} =\psi g_{\alpha \beta},
\label{diventropiaIIIa}
\end{equation}
for an arbitrary function $\psi$, then 
\begin{equation}
S^{\alpha}_{;\alpha} =0.
\label{diventropiaIII}
\end{equation}

 However we know that LTB spacetimes do no admit CKV \cite{53}, accordingly at least in the context of the standard irreversible thermodynamics, our tilted observer detects a real dissipative process.

Let us now consider the situation for the causal thermodynamics. In this latter case we obtain from (\ref{diventropia})
\begin{equation}
S^{\alpha}_{;\alpha} = -\frac{1}{2} T^{\alpha \beta}_{dis.}
\mathcal{L}_\chi g_{\alpha \beta} -\frac{1}{2}\left(\frac{q^2V^\mu \tau}{\kappa T^2}\right)_{;\mu}.
\label{diventropiaIV}
\end{equation}
The equation above allows two possibilities for the vanishing of $S^{\alpha}_{;\alpha}$. Either the two terms on the right cancel each other  or  both terms vanish separately. Now, since the second term on the right of (\ref{diventropiaIV}) contains two phenomenological parameters ($\kappa$ and $\tau$) which are absent in the first term, it follows that the vanishing of $S^{\alpha}_{;\alpha} $ in  the first case would imply a specific relationship between those two parameters. While this situation  is possible, it would refer to a specific example and is certainly not describing a generic scenario. Therefore we shall consider the second case, which requires the vanishing of both terms simultaneously. However this is not possible in our case since LTB as mentioned before does not admit CKV, and therefore we conclude that  also in the context of the Israel-Stewart theory, there is  entropy production associated with the heat flow vector $q^\alpha$.

 For spacetimes admitting CKV (e.g.  FRW) vanishing entropy production requires the vanishing of the last term in (\ref{diventropiaIV}), which  implies 
\begin{equation}
C^\mu_{;\mu}\equiv\frac{1}{\sqrt{-g}}\frac{\partial(\sqrt{-g}C^\mu)}{\partial x^\mu}=0,
\label{ld}
\end{equation}
with $g$ being the metric determinant and $C^\mu \equiv \left(\frac{q^2V^\mu \tau}{\kappa T^2}\right)$, implying in turn the conservation of $\frac{\sqrt{-g}q^2\tau}{\kappa T^2}$. At this point we do not know what is the physical meaning (if any)  of such quantity, nor can we understand why  its conservation is necessary for the reversibility of the process.

It is worth mentioning  that reversible dissipative processes may occur in collisionless plasma, an example of which is the well known Landau damping \cite{73}. In that  case, the dissipation is related to electrons whose speed in the direction of propagation of an electric wave, equals the phase speed of  the latter. It is not clear to us if there is some link between Landau damping and reversible processes satisfying (\ref{diventropiaIIIa}) and (\ref{ld})

The remarkable fact that the tilted observer  detects a real (entropy producing) dissipative process while for the non--tilted observer the evolution proceeds adiabatically, requires a deeper analysis. For doing that we shall heavily rely on a discussion presented in \cite{74} where it is shown that forces may be interpreted  in terms of collisional terms appearing in the Boltzmann equations, and thereby producing entropy. Basically, what authors of \cite{74} show is that a specific collisional interaction may be mapped onto an effective force, implying thereby that there exists a certain freedom to interpret collisional events (producing entropy) in terms of forces (and viceversa). 

Now, we have seen that for the  tilted observer the congruence of $V^\mu$ is (in general) non--geodesic, leading such an observer to conclude that some ``force'' other than gravitation is acting on the fluid. If we interpret this ``force'' as a collisonal term in the Boltzman equation we can understand why the tilted observer detects a truly dissipative process while for the non--tilted observer the evolution occurs adiabatically. In this sense it could be said that the four--acceleration is producing extra entropy. On the other hand we saw that dissipation is a distinctive characteristic associated to tilted observers, accordingly it could also be concluded that those dissipative fluxes are at the origin of the observed four--acceleration.

\section{The inhomogeneity factor and its evolution}
As it is well known, the energy--density in the ``standard'' (non--tilted) LTB is inhomogeneous. The same is true for the tilted LTB, however in this latter case the physical factors related to that inhomogeneity and their evolution are different. We shall now elaborate on this issue in some detail.

First of all observe that from (\ref{ellis1}) it follows that, assuming the fluid to be regular everywhere, 
\begin{equation}
\Psi=0 \Leftrightarrow \tilde \mu^\dagger=0,
\label{inho1}
\end{equation}
with 
\begin{equation}
\Psi \equiv X_{TF}-\frac{\int^x_0AR^3dx'}{R^3},
\label{inh2}
\end{equation}
where

\begin{equation}
A\equiv 12\pi\tilde q\left[\omega(\ln R)^\dagger+\frac{\dot R \sqrt{1-\omega^2}}{R}\right],
\label{ih3}
\end{equation}
and $x$ is a parameter of the curves of the congruence defined by $s^\alpha$.  We shall refer to $\Psi$ as the inhomogeneity factor.

In the non--tilted case ($\omega=\tilde q =0$), we have $X_{TF}=-{\cal E}$, and therefore the Weyl tensor becomes the inhomogeneity factor, a well known result \cite{75}.

Next, we shall use (\ref{ellis2}) to find the evolution of $\Psi$.  Replacing (\ref{inh2}) into (\ref{ellis2}) and integrating we obtain
\begin{widetext}
\begin{equation}
\Psi=\frac{\int^s_0\left[4\pi \tilde q^\dagger+4\pi\tilde q (\omega \Theta+\frac{2 \dot \omega}{\sqrt{1-\omega^2}}-\frac{R'\sqrt{1-\omega^2}}{BR})+4\pi \tilde \mu \sigma\right]R^3 ds' -\int^x_0AR^3dx'}{R^3},
\label{inh4}
\end{equation}
\end{widetext}
where $s$ is the parameter of the curves of the congruence of  $V^\alpha$.
In the non--tilted case (\ref{inh4}) becomes 
\begin{equation}
{\cal E}=-\frac{4 \pi \int^t_0 \mu \sigma R^3 dt}{R^3},
\label{inh5}
\end{equation}
implying that deviations from an initially homogeneous configuration depend on the shear \cite{75}. We see that in the tilted version the situation is by far more complicated, and deviations from an initially homogeneous configuration depend also on the dissipative flux.

\section{On the stability of the non--tilted congruence}
Let us consider a  non--tilted congruence, which at $t=0$ is submitted to perturbations  keeping the spherical symmetry. For simplicity we shall consider the possibility of dissipation only in the pure diffusion case ($\epsilon=0$).

We shall study the perturbed system on a time scale which is small as compared to the thermal relaxation time and the hydrostatic time scale.

Then, immediately after perturbation (``immediately'' understood in the sense above), 
\begin{equation}\omega = q = 0 ; \qquad \dot\omega \approx \dot q \not = 0 \; (small).\label{omyQ0}
\end{equation}

With the above conditions we obtain from (\ref{Bi1}) evaluated  just after the perturbation
\begin{equation}
\mu \dot \omega+\dot q=0.
\label{ins1}
\end{equation}
Next, evaluating the transport equation (\ref{treq1}) just after the perturbation, we obtain 
\begin{equation}
\tau \dot q=-\kappa T \dot \omega,
\label{ins2}
\end{equation}
where use has been made of the fact that just before the perturbation the system is geodesic and in thermal equilibrium, and therefore $T^{\prime}=0$. Combining (\ref{ins1}) with (\ref{ins2}) we obtain
\begin{equation}
\dot \omega(\mu-\frac{\kappa T}{\tau}) \equiv\dot \omega \mu (1-\alpha)=0,
\label{ins3}
\end{equation}
with $\alpha \equiv \frac{\kappa T}{\mu\tau}$.

From the above it is obvious that if $\alpha \neq 1$ then $\dot \omega=0$ and taking repeatedly, time derivative of  (\ref{Bi1}) and (\ref{treq1}) it follows that time derivatives of any order of $\omega$ vanish, implying that the non--tilted congruence can be analytically extended beyond $t=0$.

However if $\alpha=1$, it is not longer possible to assure the stability of the non--tilted congruence after perturbations. The situation described by such a condition has been studied in detail in the past, (see \cite{13}, \cite{76}--\cite{83} and references therein for details). Basically that condition (usually referred to as the ``critical point'') implies the vanishing of the effective inertial mass density and because of the equivalence principle, of the passive gravitational mass density, leading to important consequences in the dynamics of gravitational collapse.

In order to evaluate the circumstances under which such condition appear, observe that in c.g.s. units

\begin{equation}
\kappa T = \frac{G}{c^5} \, [\kappa] \, [T], \quad \tau = c \, [\tau], \qquad \mu = \frac{G}{c^2} \, [\mu],
\label{kapT}
\end{equation}
\noindent
where  $G$ is the gravitational constant  ($G \equiv 6.67\times 
10^{-8} \, g^{-1} cm^3 \, s^{-2}$) and   $[\kappa]$, $[T]$, $[\tau]$ and  $[\mu]$ denote 
the numerical value of these quantities in $erg \, s^{-1} \, cm^{-1} 
\, K^{-1}$, $K$, $s$ and $g \, cm^{-3}$ respectively.

Thus
 
\begin{equation}
\alpha\equiv \frac{\kappa T}{\tau \mu} \approx \frac{1}{81} \,  
\frac{[\kappa]\,[T]}{[\tau]\,[\mu]} \times 10^{-40}.
\label{men40}
\end{equation}

At present we may speculate that  $\alpha$ may increase substantially (for non-negligible values of $\tau$) in a pre-supernovae event. Indeed, at the last stages of massive star evolution, the decreasing of the opacity of the fluid, from very high values preventing the propagation of neutrinos (trapping \cite{84}), to smaller values, gives rise to neutrino radiative heat conduction. Under these conditions both $\kappa$ and $T$ could be sufficiently large as to imply a substantial increase of $\alpha$. In fact, the values suggested in \cite{85} ($[\kappa] \approx10^{37}$;$[T] \approx 10^{13}$; $[\tau] \approx 10^{-4}$; $[\mu] \approx10^{12}$, in c.g.s. units and Kelvin) lead to $\alpha \approx 1$. 

\section{Conclusions}
We have seen that LTB spacetimes as seen by a tilted observer exhibit physical properties which drastically differ from those present in the standard non--tilted LTB. Particular attention deserves the occurrence of dissipative fluxes which are associated to ``real'' (irreversible) dissipative processes. We put forward a qualitative explanation for the presence of such processes based on the equivalence between forces and collision terms discussed in \cite{74} and  the fact  that the congruence of $V^\mu$ is non--geodesic.

Next we have isolated the inhomogeneity factor, which differs drastically from the corresponding factor in the non--tilted case. There too the dissipative fluxes makes the difference. This result appears to be relevant with respect to  the Penrose's proposal \cite{86} to define a gravitational arrow of time. Indeed, since the rationale
behind Penrose's idea is that tidal forces tend to make the
gravitating fluid more inhomogeneous as the evolution proceeds,
thereby indicating the sense of time, it should be clear that all factors associated to energy--density inhomogenity (and not only the Weyl tensor)  should be present in any definition of the gravitational arrow of time {\it a la} Penrose, implying thereby that such definition would be also congruence dependent.

Finally we have shown that under extreme conditions (the critical point) the non--tilted congruence might be unstable, meaning that if that condition is attained, the ``natural'' version of the model would be a non--tilted one.

We would like to conclude with three remarks:
\begin{itemize}
\item It should be emphasized that our goal here is not to provide specific models for given astrophysical  scenarios, but just to bring out the relevance of the role of the observer in the description of  physical phenomena.
\item Since the physical interpretation of  both models (tilted and non--tilted) is so different, one could ask what interpretation is the better one? However we agree  with Cooley and Tupper  \cite{4}, in that the key issue is not: what the ``correct'' interpretation  of the model is? since both are physically viable. The point is that  each  interpretation is related to a specific congruence of observers, and  the subjective element  ensuing from any specific choice brings out the relevance of the observer  in the description of  a physical  phenomenon. This should not be taken as weakness of the theory but quite the opposite as  expression of its richness.
\item In a recent work \cite{53} some of us tried to generalize LTB as to admit dissipative fluxes, here we have seen that a simple way to do that is just to look at LTB from the point of view of a tilted observer.
\end{itemize}

\begin{acknowledgments}
LH wishes to thank Fundaci\'on Empresas Polar for financial support and Departamento   de F\'{\i}sica Te\'orica e Historia de la  Ciencia, Universidad del Pa\'{\i}s Vasco, for financial support and hospitality. ADP  acknowledges hospitality of the
Departamento   de F\'{\i}sica Te\'orica e Historia de la  Ciencia,
Universidad del Pa\'{\i}s Vasco. This work was partially supported by the Spanish Ministry of Science and Innovation (grant FIS2010-15492). 
\end{acknowledgments}

\end{document}